\documentclass[aps,prl,reprint,superscriptaddress,floatfix]{revtex4-1}

\usepackage{cancel}
\usepackage{amsmath}
\usepackage{amssymb}
\usepackage{graphicx}
\usepackage{xcolor}

\pdfoutput=1

\newcommand{\eq}{\begin{equation}}
\newcommand{\eeq}{\end{equation}}
\newcommand{\ket}[1]{\left|#1\right\rangle}
\newcommand{\bra}[1]{\left\langle #1\right|}

\begin{document}

\title{Coherent Imaging Spectroscopy of a Quantum Many-Body Spin System}
\author{C. Senko$^1$, J. Smith$^1$, P. Richerme$^1$, A. Lee}
\affiliation{Joint Quantum Institute, University of Maryland Department of Physics and National Institute of Standards and Technology, College Park, MD  20742}

\author{W. C. Campbell}
\affiliation{Department of Physics and Astronomy, University of California, Los Angeles, CA  90095}

\author{C. Monroe}
\affiliation{Joint Quantum Institute, University of Maryland Department of Physics and National Institute of Standards and Technology, College Park, MD  20742}


\date{\today}

\begin{abstract}
Quantum simulators, in which well controlled quantum systems are used to reproduce the dynamics of less understood ones, have the potential to explore physics that is inaccessible to modeling with classical computers. However, checking the results of such simulations will also become classically intractable as system sizes increase. In this work, we introduce and implement a coherent imaging spectroscopic technique to validate a quantum simulation, much as magnetic resonance imaging exposes structure in condensed matter. We use this method to determine the energy levels and interaction strengths of a fully-connected quantum many-body system. Additionally, we directly measure the size of the critical energy gap near a quantum phase transition. We expect this general technique to become an important verification tool for quantum simulators once experiments advance beyond proof-of-principle demonstrations and exceed the resources of conventional computers.
\end{abstract}

\maketitle

Certain classes of quantum many-body systems, which describe a variety of interesting problems including high-$T_c$ superconductors or spin liquids, are believed to be fundamentally inaccessible to classical modeling \cite{Cirac2012}. For example, an interacting spin system following the Ising model can map to NP-complete computational problems \cite{Cipra2000} and has been applied to understanding neural networks \cite{Schneidman2006} and social behavior \cite{Liu2010}, yet quickly becomes theoretically intractable due to the exponential number of possible spin configurations \cite{DiepFrustrationBook,SachdevQPTBook}.

Quantum simulations \cite{Feynman1982,Lloyd1996,Buluta2009,NaturePhysics2012}, in which well-controlled quantum objects like photons \cite{Aspuru-Guzik2012} or ultracold atoms \cite{Blatt2012,Bloch2012} are induced to emulate other quantum systems, are a promising alternative for accessing such problems. However, as these systems approach theoretically intractable physics, validating quantum simulation results will become a major challenge \cite{Hauke2012,Cirac2012}. Here we introduce a technique for performing coherent imaging spectroscopy on the Hamiltonian of an interacting many-body spin system. We use spectroscopic imaging to infer spin-spin interaction strengths and directly measure the critical energy gap near a quantum phase transition. This technique is a promising benchmarking tool for systems of 30+ spins, where classical computation begins to fail.

Ultracold atomic systems are particularly well-suited for simulating interacting spin systems, with the ability to prepare known input states, engineer tunable interaction patterns, and measure individual particles \cite{Bloch2012,Blatt2012}. Our experiment uses trapped ions to simulate chains of spin-1/2 particles subject to effective magnetic fields and long-range, inhomogenous Ising couplings generated by optical dipole forces \cite{Porras2004,Kim2009, Kim2010,Edwards2010,Islam2011,Britton2012,Islam2013,Richerme2013,Richerme2013a}. This results in an effective $N$-spin Hamiltonian (with $h=1$) 

\eq
H_{\mathrm{eff}} = \sum_{i<j} J_{i,j} \sigma_i^x \sigma_j^x + B(t) \sum_i \sigma_i^y,
\label{eq:Hamiltonian}
\eeq
where $\sigma_i^\gamma$ ($\gamma=x,y,z$) is the Pauli matrix for spin $i$ along direction $\gamma$; $J_{i,j} \sim J_0 |i-j|^{-\alpha}$ is a long-range coupling strength between spins $i$ and $j$ with $\alpha$ tunable between 0 and 3 \cite{Porras2004}; and $B(t)$ is the strength of a time-dependent transverse magnetic field. 

The ability to generate antiferromagnetic $J_{i,j}$ couplings of varying interaction range \cite{Porras2004,Deng2005,Britton2012,Islam2013} has recently attracted much interest in contexts such as studying the spread of correlations after a quench \cite{Schachenmayer2013,Hauke2013}, observing prethermalization of a quantum system \cite{Gong2013}, and directly measuring response functions \cite{Knap2013}. A protocol to measure the spin-spin couplings, which until now have only been fully characterized in systems of 3 spins \cite{Kim2009,Khromova2012}, will be an important validation tool for such experiments.

\begin{figure*}
\includegraphics[scale=0.85]{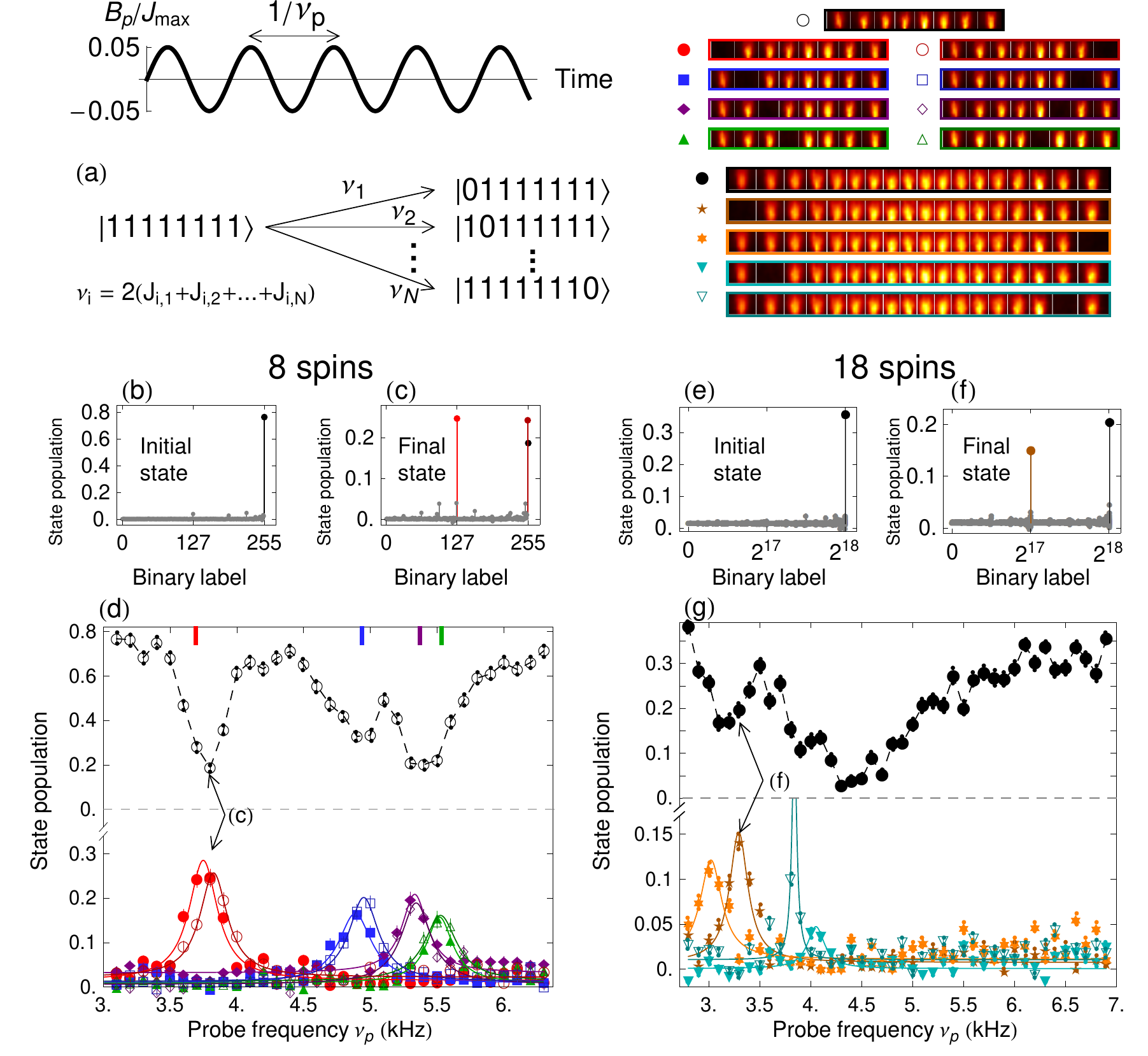}
\caption{ (a): Schematic illustration of coherent imaging spectroscopy: the transverse field is amplitude modulated with a defined frequency, driving transitions between states when the frequency matches an allowed energy splitting (e.g., between the state with all spins up along $x$ and any single-defect state with a single spin down). (b), (c): Populations of each of the $2^N$ spin states in an experiment where we prepare $N=8$ spins in state $\ket{11111111} \equiv 255$ (b), then apply a modulated field of appropriate frequency for 3 ms, producing the state shown in (c). Population is mainly transferred to the two states coupled by the probe frequency, $\ket{01111111} \equiv 127$ and $\ket{11111110} \equiv 254$ (highlighted in red; colors correspond to legend). (d): Populations in specific spin states vs. the frequency of the probe field, for a system of 8 spins.  The states are indicated in the legend of images at top right, where a fluorescing (dark) ion represents $\ket{1}$ ($\ket{0}$).  Dashed lines guide the eye; solid curves are Lorentzian fits to the data sets. Individual state imaging allows each energy splitting to be mapped separately. Colored bars indicate calculated predictions of energy splittings, and error bars show the statistical error from performing 1000 repetitions of each experiment. (e), (f): Populations in each of the $2^N$ spin states for a similar experiment with $N=18$ spins.  (g): Populations in four specific spin states vs. the frequency of the probe field, for a system of 18 spins. There is a left-right asymmetry, believed to be caused by slight misalignment of the laser beams. Despite the low fidelity of the initial state (starting with 35\% in the $\ket{111111111111111111}$ state), these energy splittings are still clearly visible. }

\label{fig:spectroscopy1flip}
\end{figure*}

The spin-1/2 particles are represented by a string of $^{171}$Yb$^+$ ions confined in a macroscopic Paul trap. The spin states $\ket{\downarrow}_z$ and $\ket{\uparrow}_z$ are encoded in the magnetic-field-insensitive ($m_F=0$) hyperfine states of the ground electronic manifold \cite{Olmschenk2007}. The spin-spin couplings and effective magnetic fields derive from lasers that globally illuminate the ion chain, driving stimulated Raman transitions between the spin states \cite{Kim2009} (see Supplementary Information). State initialization is accomplished via optical pumping into the $\ket{\downarrow\downarrow\downarrow\cdots}_z$ state followed by a coherent rotation to polarize all spins along the desired axis. After applying the spin-spin couplings and the probe field(s) described above, the individual spin states are read out along any axis by performing a coherent rotation from the desired axis to the measurement basis basis $\ket{\downarrow}_z$ and $\ket{\uparrow}_z$, then collecting state-dependent fluorescence onto a CCD imager with site-resolving optics. 

\subsection{Coherent Imaging Spectroscopy}

We measure the energy splittings in our spin system using a weakly modulated transverse field as a probe,
\eq
B(t) = B_0 + B_p \sin(2\pi \nu_{p} t).
\eeq
When the probe frequency $\nu_{p}$ is matched to the energy difference $|E_a - E_b|$ between two eigenstates $\ket{a}$ and $\ket{b}$, the field will drive transitions between the two states if there is a nonzero matrix element $\bra{b} B(t) \sum_i \sigma_i^y \ket{a} \neq 0$. For example, in the weak-field regime $B(t)\ll J_0$, the eigenstates of the Hamiltonian are symmetric combinations of the $\sigma^x$ eigenstates, and the matrix element $\bra{b} B(t) \sum_i \sigma_i^y \ket{a}$ is nonzero only when $\ket{a}$ and $\ket{b}$ differ by the orientation of exactly one spin.

In the weak-field regime, a transition at a single frequency can easily be monitored, and its stability can provide a good proxy for the entire Hamiltonian. Each splitting depends on multiple spin-spin couplings -- for example, a transition from $\ket{1111\cdots}$ to $\ket{0111\cdots}$, where $\ket{1}$ ($\ket{0}$) denotes the $\sigma^x$ eigenstate $\ket{\uparrow}_x$ ($\ket{\downarrow}_x$), requires energy 

\begin{equation}
\label{eqn:DeltaE}
\Delta E = 2(J_{1,2} + J_{1,3} + \cdots + J_{1,N})
\end{equation}

These splittings are therefore sensitive to changes in the motional mode structure or the laser intensities at each of the ions.

We demonstrate the mapping of individual energy splittings in the weak-field regime $B(t)/J_0\ll 1$ in Figure \ref{fig:spectroscopy1flip}. The spins are prepared along the $x$ direction in $\ket{111\cdots}$ and a probe field of $B(t) = (100$ Hz)$\sin (2\pi \nu_{p} t)$ is applied for 3 ms, which is sufficient to transfer more than 50\% of the population between states, before measuring along $x$. These parameters allow resolution of the energy differences in an 8 spin system while still accommodating the few ms decoherence timescale in our system \cite{Islam2013}.

\begin{figure}
\includegraphics[scale=0.6]{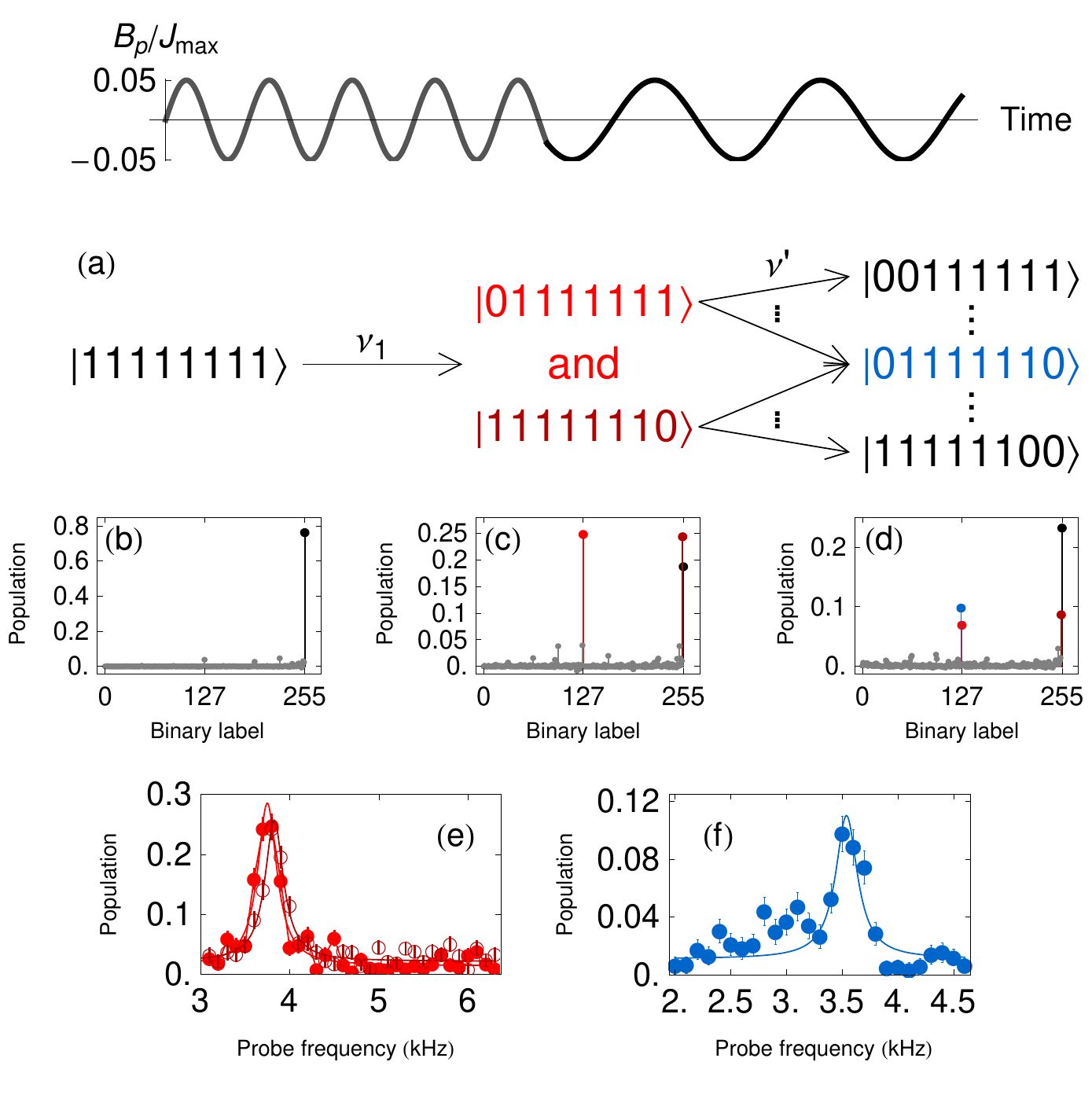}
\caption{ (a): Sketch of protocol for driving sequential excitations. (b)-(d): plots of the measured populations in each of the $2^N$ states in a system of $N=8$ spins. We apply sequential pulses of the modulated transverse field to an initial polarized state, shown in (b).  The first pulse drives transitions into states with single defects, $\ket{01111111}$ and $\ket{11111110}$ (c), and the second pulse can then create states with two defects (d). (e): Population in either of the states with a single defect on the end vs. the frequency of the first pulse.  (f): Fixing the first pulse on resonance from (e), we show the population of a state with two defects, $\ket{01111110}$, vs. the frequency of the second pulse. }

\label{fig:multipleExcitations}
\end{figure}

Population transfer is clearly seen when $\nu_{p}$ is resonant with an energy splitting. We quantify the energy of a particular state relative to the initial state by fitting the spectra to Lorentzians (see Supplementary Information). The spectral positions are insensitive to measurement error and loss of population in the initial state, which affects only the contrast of these resonances, as we show in Figure \ref{fig:spectroscopy1flip}(g) with $N=18$ spins.

A sequence of multiple probe frequencies (shown in Figure \ref{fig:multipleExcitations}) can be used to populate any desired spin configuration with a global beam in no more than $\lfloor N/2 \rfloor$ pulses. We have demonstrated the ability to transfer population into any of the 32 eigenstates of a 5 spin system by starting in either the $\ket{11111}$ or $\ket{00000}$ and applying at most two pulses of the transverse field. This system is small enough to also measure the entire relative energy spectrum, which scales exponentially with system size. Starting from the states $\ket{11111}$, $\ket{00000}$, $\ket{10101}$, and $\ket{01010}$ (the last two of which are prepared using an adiabatic ramp of a transverse field \cite{Islam2013}), we use single and multiple frequency drives to measure all possible energy splittings.

\begin{figure}
\includegraphics[scale=0.5]{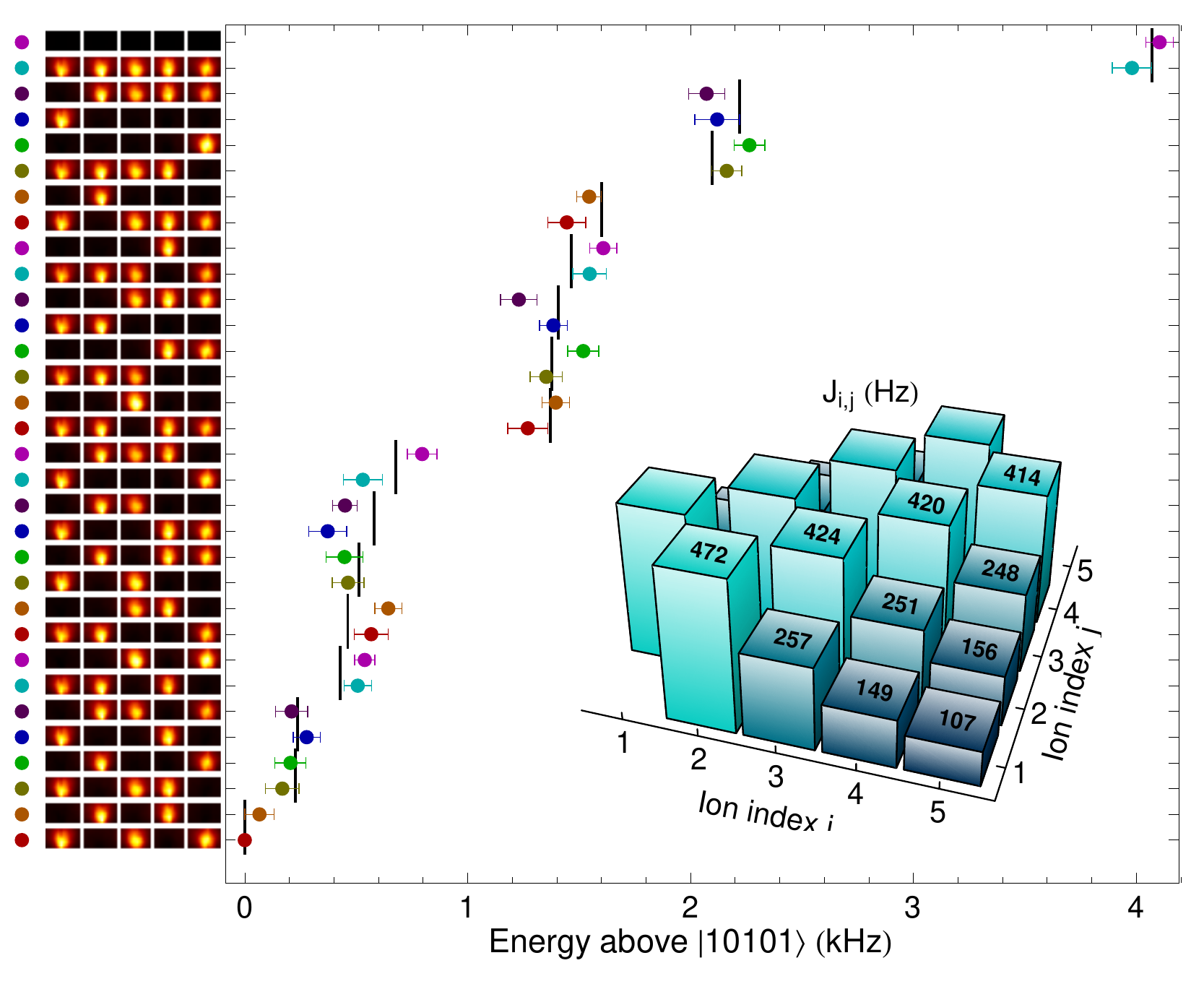}
\caption{ Reconstructed energy of each spin configuration above the $\ket{10101}$ (ground) state (colored points) compared to the calculated energies (black lines). Calculations are based on the spin-spin couplings estimated from the same energy measurements (shown in the inset). Error bars include statistical errors and an estimate of systematic error due to experimental drifts.  }
\label{fig:fullspectrum}
\end{figure}

Figure \ref{fig:fullspectrum} shows the measured spectrum of this 5-spin system, obtained by direct addition of the measured energy splittings, compared to that given by the interactions estimated from the same data (as detailed below). An examination of the full spectrum of a many-body quantum system is generally difficult to achieve, and shows the versatility of this form of many-body spectroscopy. 

\begin{table*}[t]
\begin{ruledtabular}
\begin{tabular}{| c | c | c | c | c | c | c |}
\hline
Trace over 1? & Trace over 2? & Trace over 3? & Trace over 4? & $\left< W_{ss} \right>$ & Number ions involved, N & Min. possible value of $\left< W_{ss} \right>$ \\ \hline
No & No & No & No & -1.62(22) & 4 & -3 \\ \hline
Yes & No & No & No & -0.382(121) & 3 & -2 \\
No & Yes & No & No & -0.847(96) & 3 & -2 \\
No & No & Yes & No & -0.735(101) & 3 & -2 \\
No & No & No & Yes & -0.300(114) & 3 & -2 \\ \hline
Yes & Yes & No & No & -0.115(40) & 2 & -1 \\
Yes & No & Yes & No & -0.111(41) & 2 & -1 \\
Yes & No & No & Yes & -0.001(44) & 2 & -1 \\
No & Yes & Yes & No & -0.279(37) & 2 & -1 \\
No & Yes & No & Yes & -0.081(38) & 2 & -1 \\
No & No & Yes & Yes & -0.055(39) & 2 & -1 \\ \hline

\end{tabular}
\end{ruledtabular}
\caption{ Measured values of the spin-squeezing-type witness $W_{ss}$ described in the text, compared to theoretical values for a perfect 4-spin state $\ket{\Psi_W}$; a negative value certifies that the state is nonseparable and hence that at least two of the spins are entangled. By tracing over individual spins, we see that all pairs except ions 2 and 3 are at least 1$\sigma$ below zero, showing that these pairs are entangled; entanglement between each possible pair is consistent with the multipartite entanglement that would be expected for a perfect W state. }
\end{table*}

\subsection{Engineering Coherent Quantum States}
We can also use modulated transverse fields to prepare arbitrary coherent quantum states, which can be used to probe many-body quantum dynamics \cite{Gong2013}. In general, subjecting a left-right symmetric state to a global resonant probe field prepares symmetric superpositions of states (such as those shown in Figure \ref{fig:spectroscopy1flip}(c)) that exhibit a degree of entanglement, though this is difficult to detect without individual rotations for readout.

We generate entanglement by subjecting an initial state $\ket{111\cdots11}$ to multiple frequencies simultaneously, such that all of the possible transitions are driven equally. After an appropriate time, the system will ideally be driven into a W-type state of the form 
\begin{eqnarray}
\ket{\Psi_{W}} =& \frac{1}{\sqrt{N}} \left  (e^{i\phi_0}\ket{011\cdots11}+e^{i\phi_1}\ket{101\cdots11}+\right. \\
\nonumber
&\left. \cdots+e^{i\phi_1}\ket{111\cdots01}+e^{i\phi_0}\ket{111\cdots10} \right ),
\label{W}
\end{eqnarray} 
where the phases $\phi_i$ depend on the relative phase of the applied modulation frequencies. Entanglement can then be detected using global measurements of the magnetization along various directions of the Bloch sphere \cite{Toth2007}. In particular, we use a witness operator
\eq
W_{ss} = (N-1) (\left<J_x^2\right> - \left<J_x\right>^2) + \frac{N}{2} - \left<J_y^2\right> - \left<J_z^2\right>,
\eeq
where $J_\gamma \equiv \frac{1}{2} \sum_{i=1}^N \sigma_i^\gamma$ (with appropriate phases) and angle brackets denote ensemble averages. The significance of this spin-squeezing observable is that it will always be positive for separable states, so measurement of a negative value certifies that at least two particles are entangled.

We prepare an entangled state of 4 spins by applying two simultaneous frequencies of the modulated transverse field to the state $\ket{1111}$ with an appropriate relative phase for 1.8 ms and measure the resulting state along the Bloch sphere directions $x$, $y$, and $z$ to obtain the witness shown above. Moreover, individual spin state imaging allows us to trace over any given spin or pair of spins and apply the witness to this reduced density matrix. The data is shown in Table 1, which certifies that the full state as well as every possible reduced state are entangled.

\begin{figure}
\includegraphics[width=\columnwidth]{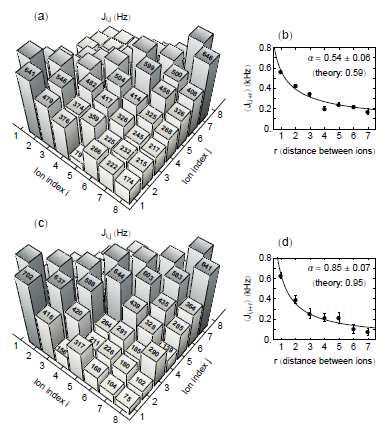}
\caption{ Two experimentally determined spin-spin coupling profiles in a system of eight spins. The couplings were measured for two sets of trap parameters, corresponding to a more long-range (top) or more short-range (bottom) interaction profile. (a) and (c) depict the individual elements of the measured coupling matrix. (b) and (d) plot measured average nearest-neighbor, average next-nearest-neighbor, etc., interactions against ion separation and show a fit to a power law $J_0/r^\alpha$. The error in $\alpha$ is an estimate of the standard error in the fit parameter; this takes into account the errors in the $J_{i,j}$ estimates (not shown) based on fit error and statistical error in population measurements. Further details of error analysis are in Supplementary Information. }

\label{fig:couplingmatrix}
\end{figure}

\subsection{Verifying many-body interactions}
Many-body spectroscopy using a transverse probe field further enables determination of each individual spin-spin coupling $J_{i,j}$. Using only $N+1$ scans of the probe frequency, we can measure ${N \choose 2} = N(N-1)/2$ energy splittings and thus determine the entire interaction matrix of $N \choose 2$ couplings (e.g. Eq. \ref{eqn:DeltaE}).  For example, one scan probes the state $\ket{1111\cdots}$ and yields the $N$ energy splittings to the single-defect states. Then, as in Figure \ref{fig:multipleExcitations}, $N$ additional scans starting from each single defect state determine $N-1$ further energy splittings.  In total, these $N+1$ scans yield $N^2$ measurements ($N$ from the first probe scan and $N(N-1)$ from the rest). Due to the parallel processing enabled by imaging individual spin states, the number of measurements to evaluate all spin-spin couplings scales only linearly with the system size.

We perform this verification protocol on a system of 8 spins with two different interaction ranges. We can measure the full interaction matrix with 5 frequency scans: due to the left-right symmetry, single-defect states are populated in pairs and only 4 scans are necessary to probe all 8 of the defect states. The obtained matrix agrees well with theory; roughly 70\% of measured interactions match the prediction within 1$\sigma$ standard error. As shown in Figure \ref{fig:couplingmatrix}, we observe a clear distinction in the coupling matrices for differently chosen ranges of spin-spin interactions.

\subsection{Measuring a critical gap}

\begin{figure}[t]
\includegraphics[width=\columnwidth]{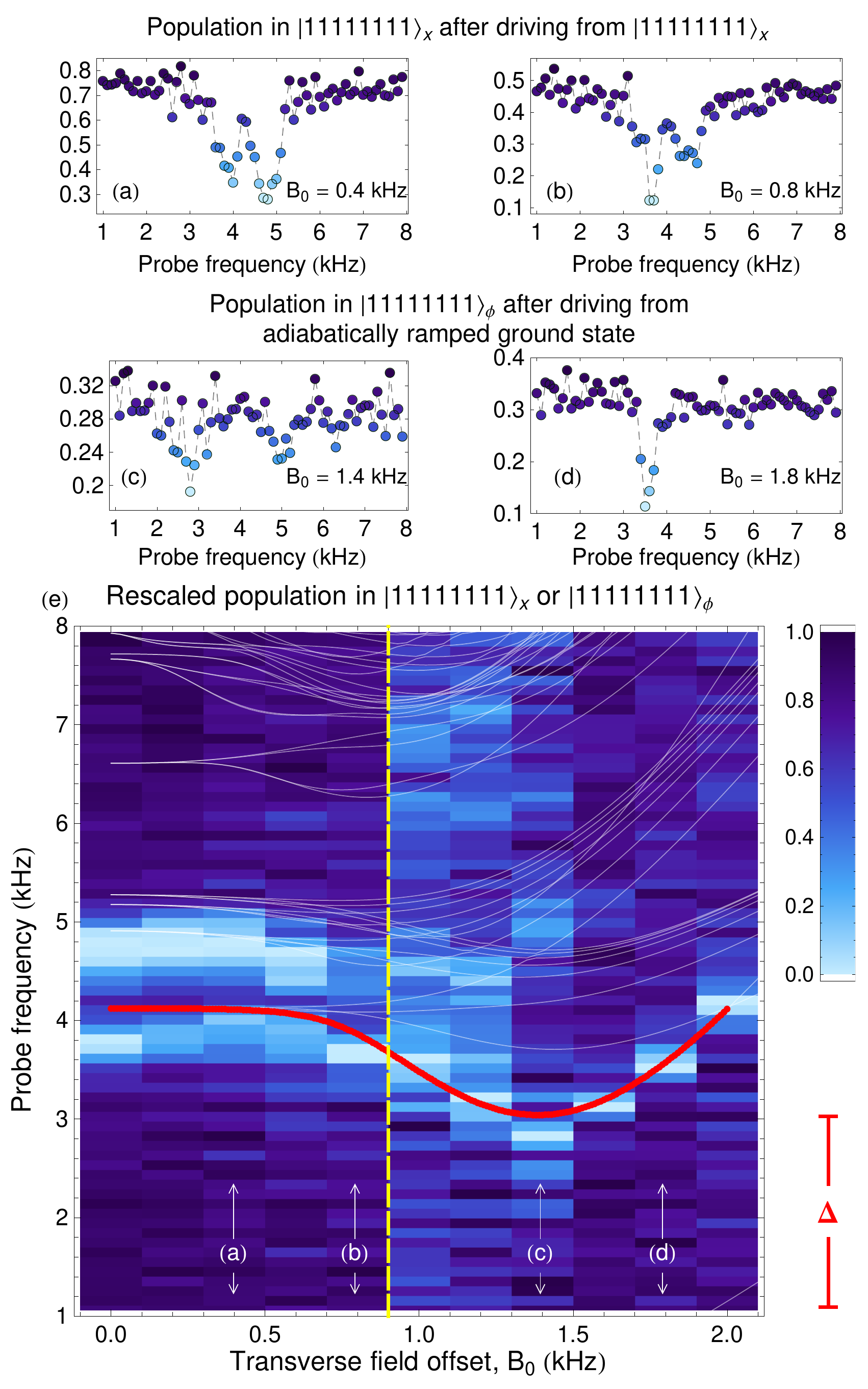}
\caption{ (a)-(d): Populations in a polarized state vs modulation frequency of the transverse field at four different values of the offset field $B_0$. Coloring is according to the rescaling scheme used in (e). In (a) and (b), we subject the state $\ket{11111111}$ to the modulated field, then measure its population. In (c) and (d), we prepare the ground state via an adiabatic ramp, subject it to the modulated field, and then measure the population in $\ket{\uparrow\uparrow\uparrow\uparrow\uparrow\uparrow\uparrow\uparrow}_\phi$ (see text). 
(e): Rescaled populations in $\ket{11111111}$ (left of the dashed line) or $\ket{\uparrow\uparrow\uparrow\uparrow\uparrow\uparrow\uparrow\uparrow}_\phi$ (right of the dashed line) vs. static field offset $B_0$ and modulation frequency. Calculated energy levels, based on measurements of trap and laser parameters, are overlaid as thin white lines, and the lowest coupled excited state as a thick red line, showing the critical gap $\Delta$ at position (c). The energy of the ground state is always taken to be zero. }

\label{fig:gap}
\end{figure}

Finally, we probe energy levels at nonzero transverse field $B_0$, including near the critical region $B_0 \approx \left<J\right>$. Determining the critical energy gap $\Delta$, at which the energy difference between the ground and lowest coupled excited states is minimized, is useful because this parameter determines the ability to perform an adiabatic sweep of the transverse field \cite{Edwards2010,Richerme2013}. However, measuring the critical gap is difficult in general due to the inability of measuring in or even knowing the instantaneous eigenbasis.

The protocol described in Figure \ref{fig:spectroscopy1flip} is effective even when there is a small DC field $B_0$ (Figure \ref{fig:gap}(a) and (b)), but breaks down near the critical region.  However, for a finite-size ferromagnetic system, measurements along a different axis of the Bloch sphere (here, $\hat{x} + \hat{y}$) allow us to still observe transitions from the ground to the first coupled excited state near the critical gap (Figure \ref{fig:gap}(c) and (d)). As shown in Figure \ref{fig:gap}(e), these experiments allow us to map the lowest coupled excited state from $B_0=0$ beyond the critical energy gap $\Delta$. The downward drift in energies near $B_0=0$ can be attributed to drifts in laser and trap parameters as the experiments progressed from higher to lower fields. An alternative protocol, which follows the time evolution after a quench, has recently been proposed for measuring the critical gap and may scale better for larger systems \cite{Yoshimura}.

\subsection{Outlook}
Due to the ability to combine coherent frequency measurements with spatial imaging of the spin ordering, this spectroscopy protocol has proven to be a powerful tool for measuring parameters such as individual spin-spin couplings and critical energy gaps. The technique will no doubt benefit from further refinements borrowing from the extensive literature of spectroscopic methods developed in other fields such as NMR \cite{ErnstNMR1987}.  We anticipate that such extensions may become widely applicable to studies of many-body effects. In addition, the protocol developed here is general and will impact experiments with ion traps and other platforms as system sizes increase, both in full calibrations of the coupling matrix and in the ability to observe a single quantity that serves as a proxy for the entire Hamiltonian. Because of the importance of validation techniques for quantum simulation of classically intractable physics, our work represents an advance in the quest to scale up quantum simulators toward this goal.

\begin{acknowledgments}
We thank Jim Freericks, Bryce Yoshimura, Emily Edwards, Zhe-Xuan Gong, Michael Foss-Feig, and Alexey Gorshkov for helpful discussions. This work is supported by the U.S. Army Research Office (ARO) Award No. W911NF0710576 with funds from the DARPA Optical Lattice Emulator Program, ARO Award No. W911NF0410234 with funds from the IARPA MQCO Program, and the NSF Physics Frontier Center at JQI.
\end{acknowledgments}

\bibliography{../Bibliography/lotsofrefs}

\section{Supplementary Information}

\subsection{Spin-spin couplings}
The spin-spin couplings described in Eq. 1 result from applying a spin-dependent optical dipole force in the $\sigma^x$ basis with the Raman lasers. The wavevector difference $\Delta k$ of the two laser beam paths is along one of the principal transverse axes of the trap, allowing the lasers to couple to the collective motional modes and virtually excite phonons that mediate the spin-spin interaction. A pair of beat frequencies are symmetrically detuned from the resonant transition between the spin states, at $\nu_0 / 2\pi = $ 12.642819 GHz, by an amount $\mu$ which is of the same order as the motional mode frequencies. In the Lamb-Dicke regime, this generates a M\o lmer-S\o rensen-type interaction \cite{Sorensen2000} given by
\eq
H_{MS} = \sum_{i,m=1}^N \eta_{i,m} \Omega_i \sin (\mu t) \sigma_i^x \left[ \hat{a}_m e^{-i\omega_m t} + \hat{a}_m^\dagger e^{i \omega_m t}  \right].
\eeq
Here, $i$ and $m$ index the ions and motional modes, respectively; $\Omega_i$ represents the equivalent resonant carrier Rabi frequency at ion $i$, $\omega_m$ the frequency of the $m$th collective mode of motion, $\hat{a}_m$ ($\hat{a}^\dagger_m$) the annihilation (creation) operator for mode $m$, $\eta_{i,m} = b_{i,m}\Delta k \sqrt{\frac{\hbar}{2M \omega_m}}$ is the Lamb-Dicke factor, couping ion $i$ to mode $m$, $b_{i,m}$ is the normal mode matrix and $M$ is the mass of a single $^{171}Yb$ ion \cite{James1998}.

In the limit of a large detuning, $|\mu-\omega_m| \gg \eta_{i,m} \Omega$, the motion is only virtually excited and the effective Hamiltonian is the spin-spin interaction,
\eq
H = \sum_{i<j} J_{i,j} \sigma_i^x \sigma_j^x.
\eeq
Here the spin-spin couplings are given by \cite{Kim2009}
\eq
J_{i,j} = \Omega_i \Omega_j \Omega_R \sum_{m=1}^N \frac{b_{i,m} b_{j,m}}{\mu^2-\omega_m^2},
\eeq
where $\Omega_R = \frac{\hbar (\Delta k)^2}{2M}$ is the recoil frequency. We note that for the experiments reported here, the detuning from the center-of-mass mode $\mu-\omega_1$ was between $3\eta \Omega$ and $4\eta \Omega$, where $\eta$ is the Lamb-Dicke factor for the COM mode (which is the highest in frequency of the transverse modes), and $\Omega$ is assumed to be uniform for each ion. Thus, excitation of the COM mode is less than 10\% (and excitation of the other modes is even lower). 

The frequencies and eigenvectors of the transverse motional modes used above can be fully characterized (in the limit where the trapping potential can be well approximated by a 3-dimensional harmonic oscillator) by the secular frequencies characterizing the potential in the $x$ transverse direction and the $z$ axial direction \cite{James1998}. For the experiments reported in this work, the transverse trapping frequency is roughly $\omega_1/2\pi = $4.8 MHz and the axial trapping frequency is varied between 0.59 MHz -- 1.05 MHz.

When the beatnote is tuned to $\mu>\omega_1$, the interaction profile varies between a uniform all-to-all coupling ($J_{i,j} \sim J_0$) in the limit where the $\mu$ is close enough to $\omega_1$ to neglect contributions from other modes, to a dipolar falloff ($J_{i,j} \sim J_0/|i-j|^3$) in the limit where $\mu$ is far detuned from all the modes. In between these limits, numerical calculations show that the interaction profiles can be roughly approximated by $J_{i,j} \sim J_0/|i-j|^\alpha$ with $0<\alpha<3$. The exponent $\alpha$ depends on the relative detunings from all the modes, and can be varied either by changing $\mu$ or by changing the axial trap frequency.

In our experiments, the main causes of drifts in the Hamiltonian are due to laser intensity noise (e.g., from pointing instability), which directly affects $\Omega_i$, and to drifts in the transverse trap frequency $\omega_x^{COM}$ (e.g., due to slight internal temperature changes in the resonator delivering RF voltage to the trap), which affects all the mode frequencies $\omega_m$ and hence the detunings $\mu-\omega_m$.

\subsection{Measurement of spin states}
The detection cycle for each experiment consists of exposing the ions to `detection' light, resonant with the $\ket{\uparrow}_z$ (`bright') state but not the $\ket{\downarrow}_z$ (`dark') state, for 3 ms. An objective with a numerical aperture of NA=0.23 collects the resulting fluorescence, which is imaged onto an intensified CCD camera. To calibrate the readout, we perform 1000 cycles of preparing and measuring an all-dark state, $\ket{\downarrow\downarrow\downarrow\cdots}_z$, and 1000 cycles of an all-bright state, $\ket{\uparrow\uparrow\uparrow\cdots}_z$.  Single-shot discrimination is performed by summing the columns of the resulting image into a 1-dimensional row, since the vertical direction yields no additional information in a linear chain, and fitting the resulting profile to a sum of Gaussians whose positions and widths are determined from the calibration images. The individual ion states are then discriminated by comparing the fit amplitudes to calibrated thresholds (see below). 

The calibration also allows us to determine the detection errors for each ion, i.e. the probability of misdiagnosing a dark state as bright or vice versa for a given threshold. These known errors are used to correct the probability distributions for detection errors, while also considering standard errors from shot noise \cite{Shen2012}. 

The optimal thresholds are determined by performing a Monte Carlo simulation in which certain target states are `prepared' by randomly choosing an amplitude from the appropriate calibration ensemble (e.g., for the target state $\ket{1010\cdots}$ the amplitude of the first ion is chosen from the pool of amplitudes which were fit to the first ion in the bright calibration), discriminated with a given threshold, and corrected for the detection error given the chosen threshold. A threshold is then chosen that is insensitive to statistical fluctuations and gives corrected probability distributions that match the known input ensemble well; the recovered probability distributions are nearly identical for a wide range of threshold choices.

\subsection{Measurement of energy splittings}
In the weak-field regime, we know that the only states that are coupled are those which differ by a single spin flip. Thus, for a given input state, we know exactly which states can be reached by the modulated transverse field (e.g., from $\ket{111\cdots11}$ the only states which are coupled to first order are $\ket{011\cdots11}$, $\ket{101\cdots11}$, $\cdots$, $\ket{111\cdots10}$). For each frequency scan, we extract the population of each coupled state as a function of modulation frequency. These data sets are expected to show a peak at the actual energy splitting. 

We fit these peaks to Lorentzian functions to determine each energy splitting:
\eq
L(x,x_0,w,A,o) = A \frac{w^2}{(x-x_0)^2+w^2}+o,
\eeq
where the offset $o$ is included because in some cases the baseline population of a final state may be nonzero due to, e.g., off-resonant coupling during the first excitation pulse. The center frequency, $x_0$, thus determined is a direct measure of the energy splitting between the initial and final states. (Nearly identical results are obtained using a functional form of $A$ sech$\frac{x-x_0}{w} + o$ or $A$ sinc$^2\frac{x-x_0}{w} + o$ to fit the peaks.)
In some cases, especially when multiple sequential excitations are performed, there may be insufficient population transfer to discern a peak above the noise floor. In most cases, there will be questions of how to choose the best fit out of multiple possible fits, or whether any of the possible fits are plausible. We will now describe our method for identifying the best fit, or the absence of a good fit.

\subsubsection{Seeding the fitting routine}
We perform our fits in Mathematica using the NonlinearModelFit function. This function allows us to input a data set, a fit function, a set of weights corresponding to the measurement errors, and an initial guess for the fitting parameters, and can return parameter standard errors in addition to best fit parameters. As with most fitting routines, it is sensitive to the initial guess for the center value $x_0$, and seeding the routine with different values will return different guesses for the peak location. We therefore compare multiple fits seeded with different initial guesses. For each data set, we calculate the mean and standard deviation of the y values. Then, we select all of the points whose y value is more than 1.5 standard deviations away from the mean, and use the x values of these points as seeds to the fitting routine. 

\subsubsection{Identifying a bad fit}
Some of the fits will be immediately implausible. E.g., since the y axis is a probability, the amplitude of the Lorentzian should never exceed 1. We use several such criteria, where a fit is considered ``good'' only if it meets all of the following conditions:

\begin{itemize}
	\item $0<A<1$ eliminates those fits where an unphysical probability occurs. (Typically this is only violated when the fitter finds a local optimum in a spuriously narrow peak with an amplitude many orders of magnitude larger than 1.)
	\item $w<0.6$ kHz eliminates fits which key in on a slow variation in the background level. For a 3 ms pulse (of a strength which drives less than a $\pi$ pulse on resonance), each peak is expected to be roughly 0.15 kHz wide (based on both numerical evolution of the Schrodinger equation in our many-body system, and on the Rabi solution for driving an isolated two-level system), so this leaves a large margin for typical variation in the widths.
	\item $w>\Delta w/2$, where $\Delta w$ is the standard error on the fit parameter, eliminates a few fits which key in on two neighboring points that are slightly higher than the nearby points but still within the noise, resulting in an implausibly tall and narrow peak (which nevertheless tends to have $A<1$); such fits typically have a large $\Delta w$.
	\item $A>$ 1.5 $S$, where $S$ is the mean shot noise for the data set being fit, imposes the requirement that the signal-to-noise ratio be at least 1.5.
\end{itemize}

These criteria, as with any criteria for discriminating good fits from bad, are necessarily subjective, and are chosen to be reasonably permissive of plausible fits to noisy data while still eliminating those fits which are blatantly bad on visual inspection, e.g., they successfully eliminate all fits to those data sets which visually appear most consistent with a (noisy) flat line. Of the remaining fits, should there be multiple good fits with different $x_0$ seeds for a single data set, the fit with the highest $R^2$ is chosen for the remaining analyses.
\subsection{Measuring coupling profiles}
For each coupling profile, frequency scans are performed with initial states of $\ket{11111111}$ and all single-defect states thereof, i.e. $\ket{01111111}$, $\cdots$, $\ket{11111110}$. The single-defect states are prepared with a pulse of the modulated transverse field. Using the methods described above, we extract all of the measurable energy splittings. 
For each energy splitting, we know the initial and final spin ordering and so know how to relate the energy difference to the spin-spin couplings, as described in the main text. We use these relations to build a design matrix $\textbf{A}$ and a response vector $\vec{y}$ such that $\textbf{A} . \vec{x} = \vec{y}$, where $\vec{y}$ is a vector consisting of the measured energy splittings, $\vec{x}$ is a vector consisting of the spin-spin couplings, and $\textbf{A}$ is a matrix with rank equal to or greater than the number of independent couplings (i.e., $\textbf{A}$ has $N \choose 2$ columns corresponding to the $N \choose 2$ couplings, and \itshape at least \upshape $N \choose 2$ independent rows). Mathematica's LinearModelFit routine is then used to perform a linear least-squares analysis determining which vector $\vec{x}$ minimizes the sum of squares of residuals, $\min |\textbf{A} \vec{x} - \vec{y}|$. Additionally, the data points (i.e. energy splittings) are weighted according to $1/\sigma_{x_0}$, where $\sigma_{x_0}$ is the estimated error in the fit center $x_0$ that was used to determine each splitting. 
As a side note, we have taken advantage of the knowledge that our couplings are roughly of the form $J_0/r^\alpha$ (and in particular, all share the same sign), which means that single-defect states will always be lower energy than the polarized state in our system, and (for the systems presented here) two-defect states are lower in energy than single-defect states. Without this knowledge it would be necessary to determine not only the magnitude of the energy splittings, as we do here, but also the sign, in order to fully constrain the coupling matrix. This is however achievable by making a second set of measurements with a known longitudinal field $B_x \sum_{i=1}^N \sigma_i^x$, which will shift the energies in a known direction. Comparison between the two data sets to determine whether $B_x$ shifts the levels closer together or further apart would yield the sign of the energy splitting.

\end{document}